\documentclass[aps,prl,twocolumn,showpacs,amsmath,superscriptaddress,groupedaddress]{revtex4-1}
\usepackage{graphics}
\usepackage{epsfig}
\usepackage[colorlinks=true,citecolor=blue,linkcolor=red,linktocpage=true,pagebackref=false]{hyperref}
\usepackage{color,soul} 

\begin{document}
   \title{
Floquet scattering matrix theory of heat fluctuations in dynamical quantum conductors
         }
\author{Michael Moskalets}
\email{michael.moskalets@gmail.com}
\affiliation{Department of Metal and Semiconductor Physics, NTU ``Kharkiv Polytechnic Institute", 61002 Kharkiv, Ukraine}

\date\today
\begin{abstract}
I present the Floquet scattering matrix theory of low-frequency heat fluctuations in driven quantum-coherent conductors in the linear response regime and beyond. 
The Floquet theory elucidates the use of the Callen-Welton fluctuation-
dissipation theorem for description of heat fluctuations in a multi-terminal case.  
The intrinsic  fluctuations of energy of dynamically excited electrons are identified as the fundamental source of a heat noise not revealed by the electrical noise. 
The role of backscattering in the increase of a heat noise above the level defined by the Callen-Welton theorem, is highlighted. 
\end{abstract}
\maketitle

Recent breakthrough experiments\cite{Lee:2013fc,Jezouin:2013fx} on heat transport and dissipation in nano-scale and atomic-scale electrical conductors opens an avenue for studying an energy transport on the level where the use of quantum laws becomes mandatory.   
According to Quantum Mechanics the energy exchange  between a dynamical driving force and a driven system is probabilistic. 
Therefore, the dissipated energy is an inherently fluctuating quantity. 
This becomes especially evident in the case when the particle flux does not fluctuate.\cite{Battista:2013ew} 
The famous phrase of Rolf Landauer "The noise is the signal"\cite{Landauer:1998jh} is as actual for a heat noise as it is for a charge noise\cite{Blanter:2000wi}.  

According to the famous Callen-Welton fluctuation-dissipation theorem\cite{Callen:1951wg}  the rate of the heat, $J _{Q}$, dissipated in the driven system in the linear response regime is controlled by the equilibrium fluctuations of the flux corresponding to a driving force.
Another conclusion is that the mean squared heat dissipated per unit time, ${\cal S} _{Q}(0)$, is also controlled by the equilibrium fluctuations of the same flux. 
In the case of a conductor driven by a time-dependent voltage the corresponding flux is an electrical current. 
Thus, if a small periodic voltage $V(t) = V(t + {\cal T})$ is applied across a macroscopic conductor, the Callen-Welton theorem gives for each Fourier harmonics with amplitude $V _{ \Omega} $,
(see, e.g.  Ref.~\onlinecite{PL9})
\begin{subequations}
\label{main0}
\begin{eqnarray}
J _{Q}
=
\frac{|V _{ \Omega}| ^{2} }{2 } \,  
{\cal S} _{I} ^{eq}(\Omega)
\frac{1 }{ 
\hbar \Omega \coth\left( \frac{ \hbar \Omega }{2 k_B\theta } \right)
}
,
\label{cw}
\end{eqnarray}
\begin{eqnarray}
{\cal S} _{Q}(0)
= 
\frac{|V _{ \Omega}| ^{2} }{2 } \,  
{\cal S} _{I} ^{eq}(\Omega)
,
\label{main}
\end{eqnarray}
where ${\cal S} _{I} ^{eq}(\Omega)$ is the spectral power of fluctuations of the electrical current $I$ in the conductor in equilibrium, i.e., at $V(t) =0$. 
The superscript $"eq"$ marks quantities calculated in equilibrium and the subscripts $"Q"$ and $"I"$ mark thermal and electrical quantities, respectively. 
Combining two equations above we arrive at the following relation between the dissipated heat and its fluctuations valid in the linear response regime, 
\begin{eqnarray}
J _{Q}
=
{\cal S} _{Q} (0)
\frac{1 }{ 
\hbar \Omega \coth\left( \frac{ \hbar \Omega }{2 k_B\theta } \right)
}
.
\label{main1}
\end{eqnarray}
\end{subequations}
Note that Eq.~(\ref{main}) does not refer to an electron temperature explicitly.
This is an advantage in the low temperature limit, where an electron temperature often is difficult to measure.\cite{Giazotto:2006bo}
Notice also that this equation relates the zero-frequency heat fluctuations in driven system to the finite-frequency equilibrium fluctuations of the electrical current. 
In Ref.~\onlinecite{Kindermann:2004im}, however, the heat fluctuations beyond linear response were related to the  fourth cumulant of the non-equilibrium electrical current flowing in the system. 
The similar relation was also discussed in Ref.~\onlinecite{Battista:2013ew}

The Callen-Welton theorem is applicable if the effect of the driving force can be accounted for by an extra term in the Hamiltonian.
The importance of this requirement was highlighted in a recent paper by Averin and Pekola.\cite{Averin:2010kp}.  
These authors were concerned about the (non-)applicability of the Callen-Welton theorem to calculations of the heat conductance at finite frequencies. 
The reason is that the temperature difference, inducing a heat current, cannot be described in terms of the Hamiltonian directly. 
However, the heat current can be induced not only by a temperature difference. 
Generally speaking, the heat current can be induced by any dynamical force acting onto the system. 
This is a case, which is addressed in the present paper. 

The heat fluctuations\cite{Krive:2001ci, Averin:2010kp,Sergi:2011eo,Zhan:2011ea,Battista:2013ew} and the full counting statistics\cite{Kindermann:2004im,Golubev:2013ct} of heat dissipated in stationary conductors were already addressed in the literature. 
Here I address heat fluctuations in driven coherent conductors. 
Note that in the linear response regime the theory developed below reproduces Eqs.~(\ref{main0}), as it should.
  
To consider this problem I take advantage of the Floquet scattering matrix approach.\cite{Moskalets:2011cw} 
The model used is the following. 
The conductor is connected to $N_r$ contacts via one-dimensional leads. 
Electrons preserve phase coherence while propagating through conductor from one contact to the other one.  
The contact $ \alpha$ is considered as the source of equilibrium non-interacting electrons characterising by the Fermi distribution functions, $f _{ \alpha}(E) = \left\{ 1 + \exp\left( [E - \mu _{ \alpha}]/[ k_B \theta _{ \alpha}] \right) \right\} ^{-1}$  with the temperature $ \theta _{ \alpha}$ and the electrochemical potential $ \mu _{ \alpha} = \mu _{F} + eV _{ \alpha}$ with $\mu _{F}$ being the Fermi energy the same for all the reservoirs, $V _{ \alpha}$ being a stationary electric potential possibly applied to the reservoir $ \alpha = 1, \dots, N_r$. 
For simplicity we assume electrons to be spin-less fermions.   

The basic quantity is the operator for a heat current, $\hat J _{Q, \alpha}$, entering the contact $ \alpha$ (the positive direction is from the conductor to the contact),\cite{Battista:2013ew,Ludovico:2013uw} 
\begin{eqnarray}
\hat{J} _{Q,\alpha}(t) & = & \frac{1}{h} \iint_{0}^{\infty} dE dE^\prime  \left( \dfrac{E + E^\prime}{2} - \mu _{ \alpha} \right) e^{i \frac{E - E^\prime}{\hbar} t} 
\label{05} \\
&& \times \left\{ b_{\alpha}^\dag(E) b_{\alpha}(E^\prime) - a_{\alpha}^\dag(E) a_{\alpha}(E^\prime) \right\} 
. 
\nonumber
\end{eqnarray}
The operator $\hat a _{ \alpha}$ is for electrons coming from the reservoir $ \alpha$, while the operator $\hat b _{ \alpha}$ is for electrons scattered into the reservoir $ \alpha$.\cite{Buttiker:1992vr} 
In the case of a conductor driven periodically with frequency $\Omega$, these operators are related to each other via the Floquet scattering matrix\cite{Moskalets:2002hu} 
\begin{eqnarray}
\hat b _{ \alpha}(E) = 
\sum_{n=-\infty}^{\infty} 
\sum _{\beta = 1}^{N_r} 
{\rm S} _{F,\alpha \beta} (E, E_n) \hat a _{\beta}(E_n)
\,, 
\label{06}
\end{eqnarray}
where $E _{n} = E + n \hbar \Omega$, $n$ is an integer. 
The equations (\ref{05}) and (\ref{06}) are written in the wide-band limit, i.e., when the Fermi energy of electors in contacts, $\mu _{F}$, is the largest energy scale in the problem, in particular, $ \hbar \Omega \ll \mu _{F}$.


The quantities of interest are the dc heat current, $J _{Q,\alpha} = \int_0^{\cal T} (dt/ {\cal T}) \left\langle \hat{J}_{Q,\alpha}(t) \right\rangle$, flowing into the contact $ \alpha$, and the average of a symmetrized correlation function of heat currents flowing into contacts $ \alpha$ and $ \beta$, the zero-frequency heat noise power,
\begin{eqnarray}
{\cal S} _{Q, \alpha \beta}(0) 
= 
\frac{1 }{2 }
\int _{0} ^{ {\cal T}} \frac{dt }{ {\cal T} } \int\limits d \tau 
\quad
\label{07} \\ 
\langle \Delta \hat J _{Q,\alpha}(t) \Delta \hat J _{Q,\beta}(t - \tau) + \Delta \hat J _{Q,\beta}(t - \tau)  \Delta \hat J _{Q,\alpha}(t) \rangle
,
\nonumber
\end{eqnarray}
where ${\cal T} = 2 \pi/ \Omega$ is the period of  a driving force. 
The operator of heat current fluctuations is $\Delta \hat J _{Q,\alpha}(t)  = \hat J _{Q,\alpha}(t)  - \left\langle \hat J _{Q,\alpha}(t)  \right\rangle$, where $\left\langle \dots \right\rangle$ stands for the quantum statistical average over the equilibrium state of electrons incoming from contacts. 
The average of the product of two operators is the following, $\left\langle \hat a _{ \alpha} ^{ \dag}(E_1)  \hat a _{ \beta} (E_2) \right\rangle = \delta _{ \alpha \beta}\delta(E_1 - E_2) f _{ \alpha}(E_1) $.
The average of the product of more than two operators is calculated using the well known Wick's theorem.  
The customary calculations give,\cite{{Moskalets:2004ct}}
\begin{eqnarray}
J _{Q,\alpha}     
&=&   
\int_0^\infty 
\dfrac{dE}{h} \left(E - \mu _{ \alpha} \right) 
\sum\limits_{n=-\infty}^{\infty} 
\sum\limits _{\beta = 1}^{N_r} 
\label{08a} \\
&&
\left\{ f _{ \beta}\left(E_{n} \right) - f _{ \alpha}(E) \right\} 
\left |  {\rm S} _{F,\alpha \beta}\left(E,\, E_{n} \right) \right |^2  
, 
\nonumber
\end{eqnarray}
\begin{widetext}
\begin{eqnarray}
{\cal S}_{Q,\alpha\beta}(0)  
=
\delta _{ \alpha \beta}  
\frac{  \pi \left( k_{B}\theta_ \alpha \right)^{3}  }{6 \hbar } 
+
\int_0^\infty  
\frac{dE }{h }
\sum\limits_{n=-\infty}^\infty  
\Big\{
\quad\quad\quad
\label{08b} \\
\sum\limits_{p=-\infty}^\infty   
\left(E _{n} - \mu_{\alpha} \right) 
\left(E _{p} - \mu_{\beta}  \right)
\sum\limits_{q=-\infty}^\infty  
\sum\limits_{\gamma  = 1}^{N_r } 
\sum\limits_{\delta  = 1}^{N_r } 
F_{\gamma \delta } \left( {E, E_{q}  } \right)\, 
{\rm S} _{F,\alpha \gamma }^* \left( E _{n}, E \right)\, 
{\rm S} _{F,\alpha \delta }^{} \left( E _{n} , E_{q}   \right)\, 
{\rm S} _{F,\beta \delta }^* \left( E_{p} , E_{q}   \right) 
{\rm S} _{F,\beta \gamma }^{} \left( E_{p}, E  \right)\,
\nonumber \\
-
\left(E _{n} - \mu_{\alpha}  \right) 
\left(E _{} - \mu_{\beta}  \right)
F_{\beta \beta } \left( E, E  \right)\, 
\left | 
{\rm S} _{F,\alpha \beta }^{} \left( E _{n}, E \right)
\right |^2
-
\left(E  - \mu_{\alpha}  \right) 
\left(E _{n} - \mu_{\beta}  \right)
F_{\alpha \alpha } \left( {E, E } \right)\, 
\left |  
{\rm S} _{F,\beta \alpha }^{} \left( {E_{n}, E } \right)
\right |^2
\Big\}
. 
\nonumber
\end{eqnarray}
\end{widetext}
Here $F _{ \alpha \beta}(E_{1},E_{2}) = 0.5 \{ f_ \alpha(E_{1}) [ 1 - f_ \beta(E_{2}) ]  +  f_ \beta(E_{2})  [ 1 - f_ \alpha(E_{1}) ] \}$. 
Note that $J _{Q, \alpha}$ is a flux of excess energy (over the electrochemical potential $ \mu _{ \alpha}$), which becomes an actual heat current after being dissipated deep inside the contact $ \alpha$. 

If the scatterer is stationary then in the equations above the Floquet scattering matrix is replaced by the stationary scattering matrix, ${\rm S} _{F, \alpha \beta}(E _{n},E _{m}) \to \delta _{nm} {\rm S} _{ \alpha \beta} (E _{m})$. 
In the case if the scatterer is stationary but the ac potentials are applied at the contacts we proceed as follows. 
According to the approach of Refs.~\onlinecite{Jauho:1994te,Pedersen:1998uc} a uniform in space and periodic in time potential $V _{ \alpha}(t) = V _{ \alpha}(t + {\cal T})$ (applied to the macroscopic contact $ \alpha$ but not into the transition region near the scatterer) results in  the following  phase factor, $\Upsilon_{\alpha}(t) = \exp[-\,i\,  (e/\hbar) \int^{t} dt'\, V_{\alpha}(t') ]$,  acquired by an electron wave function in the contact $ \alpha$. 
Straightforward calculations show that in this case in the equations above we have to replace, ${\rm S} _{F, \alpha \beta} \left( E _{n}, E _{q} \right) \to  \sum_{m = -\infty}^{\infty} \left(  \Upsilon_{\alpha} ^{*}\right) _{n-m} {\rm S}_{\alpha\beta}(E _{m}) \Upsilon_{\beta, m-q}$, where $\Upsilon_{\alpha,r}$ is a discrete Fourier transform of $\Upsilon_{\alpha}(t)$.

\begin{figure}[t]
\begin{center}
\includegraphics[width=80mm]{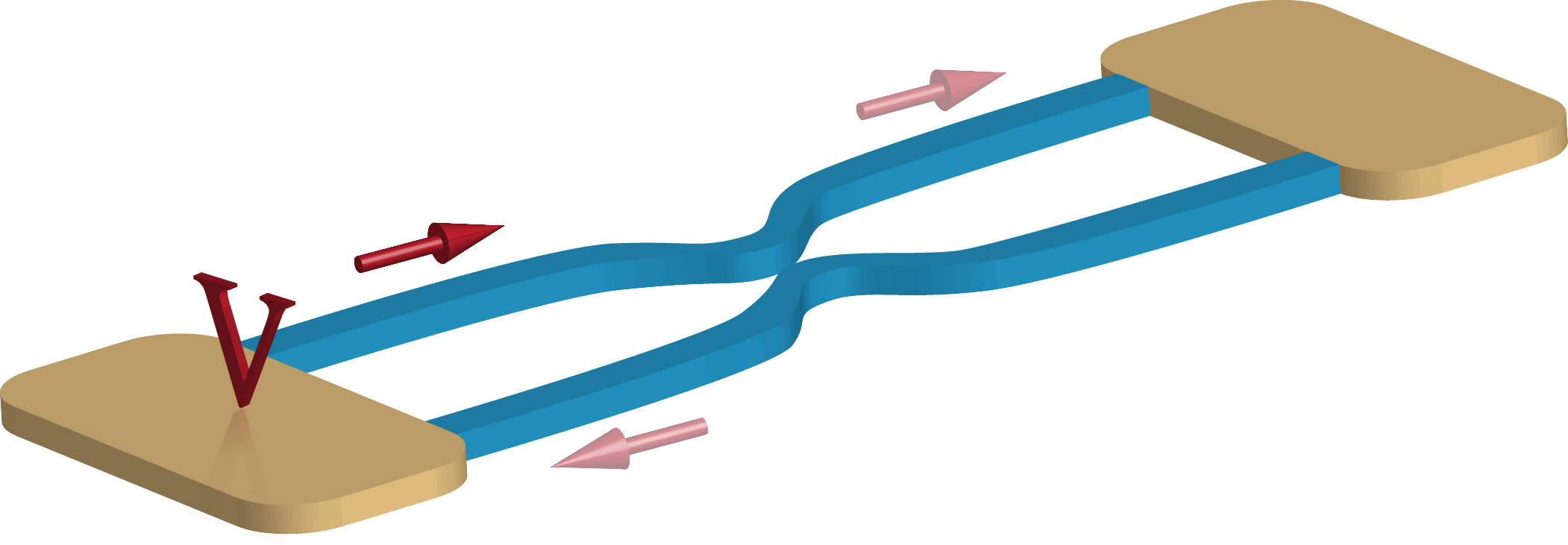}
\caption{(Color online)  A sketch of a ballistic conductor (or a Hall bar with chiral edge states) interrupted by a quantum point contact (QPC) shown as a constriction. A voltage bias $V$ across a QPC induces an electron flux, shown by arrows, transmitted and reflected at a QPC. }
\label{hallbarQPC}
\end{center}  
\end{figure}

To illustrate Eqs.~(\ref{main}) and (\ref{main1}) and to highlight main sources of the noise of a dissipated heat  let us consider simple but instructive example, namely a quantum point contact (QPC)  with energy-independent transmission, $T$, and reflection, $R = 1 -T$, probabilities, Fig.~\ref{hallbarQPC}.
The periodic voltage $V _{1}(t) = V _{10} + V _{11}(t)$ is applied to the contact $ \alpha=1$. The static part of the potential is accounted in the electrochemical potential, $ \mu _{1} = \mu + e V _{10}$, $ \mu _{2} = \mu$. 
While its ac part with zero mean is accounted for by the phase factor $\Upsilon_{1}(t)$ as described above. 
The elements of the Floquet scattering matrix read: ${\rm S} _{F, \alpha \beta} \left( E _{n}, E _{q} \right) = {\rm S}_{\alpha\beta} \left(  \Upsilon_{\alpha} ^{*} \Upsilon _{ \beta}\right) _{n-q} $ with $\Upsilon _{2} = 1$.
To ensure that the system is in equilibrium at $V _{1}(t) = 0$ we put the temperature of both contacts to be the same, $\theta _{1} = \theta _{2} = \theta$.\cite{note3}
The total heat released in the conductor, the Joule heat, is equally\cite{note1} partitioned between the contacts ($J _{Q,1} = J _{Q,2} \equiv J _{Q}$),
\begin{eqnarray}
J _{Q}
=
\frac{G}{2 } \int _{0} ^{{\cal T}} \frac{dt }{ {\cal T} }V_1^2(t) ,
\label{10}
\end{eqnarray}
where the conductance $G = T G_0$ with $G_0 = e^2/h$ being the conductance quantum. 
The correlation functions are  the following [for brevity we drop the argument  $(0)$]
\begin{widetext}
\begin{subequations}
\label{11}
\begin{eqnarray}
{\cal S} _{Q} ^{auto} &=&
{\cal S} _{Q} ^{eq}
+   
2k_B \theta 
(1 - R/ 3 )
J _{Q} 
\nonumber \\
&&
+
2k_B \theta 
\sum\limits_{q=-\infty}^\infty  
\left\{
\frac{ T^2 }{ 2h} \left | eV _{11,q} \right |^2 {\cal F}\left( \frac{q\hbar \Omega  }{2 k_B\theta } \right)
+ 
\frac{RT  }{3h } 
\left [ \left |  \left(  eV_1 \Upsilon_1 \right) _{q}  \right |^2   + \left( \pi k_B\theta \right)^2 \left | \Upsilon_{1,q}  \right |^2 \right]
{\cal F}\left( \frac{ q \hbar \Omega + e V _{10}   }{2 k_B\theta } \right)
\right\}
, 
\label{11a} \\
{\cal S} _{Q} ^{cross} &=&
- {\cal S} _{Q} ^{eq}  
+  
2k_B \theta 
\left\{  
\frac{R }{3 }  J _{Q}
+
\frac{R T }{6h }
\sum\limits_{q=-\infty}^\infty  
\left [  \left |  \left(  eV_1 \Upsilon_1 \right) _{q}  \right |^2 - 2\left( \pi k_B\theta \right)^2 \left | \Upsilon_{1,q}  \right |^2 \right]
{\cal F}\left( \frac{ q \hbar \Omega + e V _{10}   }{2 k_B\theta } \right)
\right\}
, 
\label{11b} \\
{\cal S} _{Q} &=&
2k_B \theta 
\left\{  
J _{Q} 
+
\frac{G }{2 }
\sum\limits_{q=-\infty}^\infty  
\left[ 
T
\left | V _{11,q} \right |^2 {\cal F}\left( \frac{q\hbar \Omega  }{2 k_B\theta } \right)
+
R
\left |  \left(  V_1 \Upsilon_1 \right) _{q}  \right |^2 
{\cal F}\left( \frac{ q \hbar \Omega + e V _{10}   }{2 k_B\theta } \right)
 \right]
\right\}
, 
\label{11c} 
\end{eqnarray}
\end{subequations}
\end{widetext}
where the function ${\cal F}(X) = X \coth(X) - 1$. 
To obtain the equation above I used the following identity, $\sum_{s=-\infty}^{\infty} \hbar \Omega s \Upsilon_{1,s} \Upsilon_{1,s-q}^{*}= V _{11,q}$. 
Remind that $V _{1}(t) = V _{10} + V _{11}(t)$ while $\Upsilon _{1}(t)$ depends on $V _{11}(t)$ only. 
The auto- and cross- correlation functions are ${\cal S} _{Q} ^{auto} = {\cal S} _{Q,11} =  {\cal S} _{Q,22}$ and ${\cal S} _{Q} ^{cross} = {\cal S} _{Q,12} =  {\cal S} _{Q,21} $, respectively.  
Their sum, ${\cal S} _{Q, \alpha}  = \sum _{ \beta} {\cal S} _{Q, \alpha \beta}$, is the same for both contacts in our case, ${\cal S} _{Q} \equiv {\cal S} _{Q, 1} = {\cal S} _{Q, 2}$.  
The equilibrium heat noise power across the conductor with transmission $T$ is the following: ${\cal S} _{Q} ^{eq} = T \pi \left( k_{B}\theta \right)^{3} /(3 \hbar)$.\cite{Averin:2010kp} 
Heat is not a conserved quantity since it is injected into the system by the driving force and dissipated by electrons deep inside contacts.  
As a consequence  ${\cal S} _{Q, \alpha}$ is not zero\cite{Moskalets:2004ct} unless the system is in equilibrium\cite{Sergi:2011eo}. 
Another notable difference from a charge noise is that the cross-correlation function of dissipated heat currents is not necessarily negative. 

By virtue of definition, namely the total dissipated heat, $\sum _{ \alpha} J _{Q, \alpha} = 2 J _{Q}$, and the total heat fluctuations, $\sum _{ \alpha}\sum _{ \beta} {\cal S} _{Q, \alpha \beta}  = 2 {\cal S} _{Q}$, are quantities, which are addressed in Eq.~(\ref{main1}).  
Neither the  auto- nor cross-correlation functions (with the equilibrium contribution being subtracted) do not possess this equation. 
To demonstrate this let us consider the high temperature limit, $k_B\theta \gg \hbar \Omega , e V$, where ${\cal F} \to 0$. 
In such a case we have, ${\cal S} _{Q}  = 2k_B \theta J _{Q}$.
This high-temperature result seems to be universal\cite{Jarzynski:1997uj,Crooks:1999ta,Bunin:2011kx,Averin:2011cc,Bochkov:2013bq} and is not restricted to the linear response regime addressed in Eq.~(\ref{main1}). 
The excess (marked by the superscript $"ex"$) auto- and cross-correlation functions depend  additionally on the reflection at a QPC causing the partition noise, 
${\cal S} _{Q} ^{ex,auto} = {\cal S} _{Q} ^{auto} - {\cal S} _{Q} ^{eq} =  {\cal S} _{Q} (1 - R/ 3 )$ 
and 
${\cal S} _{Q} ^{ex,cross} = {\cal S} _{Q} ^{cross} + {\cal S} _{Q} ^{eq} =  {\cal S} _{Q} R/ 3$. 
Notice the factor $1/3$, which is specific for partitioning of energy (as opposed to the factor $1$ specific for partitioning of a charge) carried by statistically uncorrelated electrons with continuous spectrum. 
The partition noise is an essentially non-equilibrium noise, which is not addressed by the Callen-Welton theorem. 
Interestingly, the ratio of excess cross- and auto-correlation functions, 
defined for $R < 1$, 
${\cal S} _{Q} ^{ex,cross}/{\cal S} _{Q} ^{ex,auto} =  R/(2 + T)$, depends neither on the temperature $\theta$ nor on the voltage bias. 
The similar ratio of electrical correlation functions is $-1$, since zero frequency auto- and cross-correlation functions of charge currents are equal to each other up to a sign.\cite{Blanter:2000wi}
The reason for that is the charge conservation. 


Let us now check Eq.~(\ref{main}) in the linear response regime: $V _{1}(t) = V _{10} + V _{11} \cos( \Omega t)$, with 
$eV _{10} \ll  k_B\theta$ and $ eV _{11} \ll \hbar \Omega$. 
In this case in Eqs.~(\ref{11}) only the harmonics with $q = 0, \pm 1$ do contribute. 
In addition we can use 
$\left(  V_1 \Upsilon_1 \right) _{\pm 1} \approx  V _{11}/2$ and 
$\left(  V_1 \Upsilon_1 \right) _{0} \approx  V _{10}$.    
With these simplifications Eq.~(\ref{11c}) gives, 
\begin{eqnarray}
{\cal S} _{Q} ^{lin} = \frac{V _{10}^2 }{2 } 2 G k_B\theta  + \frac{V _{11}^2 }{4 } G \hbar \Omega \coth\left( \frac{ \hbar \Omega }{2 k_B\theta } \right)
.
\label{12}
\end{eqnarray}
Here the superscript $"lin"$ refers to the linear response regime. 
The equations (\ref{12}) and (\ref{main}) are consistent with each other. 
When only a dc voltage is applied, $V _{10} \ne 0$, $V _{11}=0$, the heat noise power  is related to the electrical noise power at zero frequency, ${\cal S} _{I} ^{eq}(0) = 2 G k_B\theta$ (Nyquist-Johnson noise). 
The heat fluctuations and electrical fluctuations are due to thermal fluctuations of the number of electrons in the stream. 
They both vanish at zero temperature. 
In contrast, when an ac voltage is applied, $V _{10} = 0$, $V _{11} \ne0 $, the additional source of a heat noise comes into play. 
Due to probabilistic absorption of energy from the dynamical driving force, the energy of excited electrons fluctuates.
This source of a heat noise exists even at zero temperature. 
Indeed, now in Eq.~(\ref{main}) the  finite-frequency electrical noise power, ${\cal S} _{I} ^{eq}( \Omega) = G \hbar \Omega \coth\left( \hbar \Omega/ [2k_B\theta] \right)$\cite{Blanter:2000wi,note2}, has to be used, which does not vanish at zero temperature.  
Note that in this case we have contributions of two harmonics, $\Omega$ and $-\Omega$, with corresponding amplitudes entering Eq.~(\ref{main}) being $V _{\pm \Omega} = V _{11}/2$.

Note also that for a ballistic conductor, $R=0$, the equations (\ref{10}) and (\ref{11c}) do satisfy Eqs.~(\ref{main0}) (generalised to the case of many harmonics) even beyond the linear response regime.  
While the reflection at the QPC, $R>0$, increases heat noise over the level put by the Callen-Welton theorem.  
The reason is rooted in the fact that backscattered electrons interact again with the same driving potential. 
If they would be just split at the QPC and directed to contacts, which a driving potential is not applied to, then the heat noise, ${\cal S} _{Q, \alpha}$, and the  dissipated heat, $ J _{Q, \alpha}$, would satisfy the Callen-Welton theorem as they do for the case of a ballistic conductor.

To illustrate this let us consider an example,  
a periodic sequence of Lorentzian voltage pulses applied across the QPC, $eV ^{L}(t) = \sum _{n=-\infty} ^{\infty} 2 \hbar \Gamma /([t - n {\cal T}]^2 + \Gamma^2 )$, where $\Gamma \ll {\cal T}$ is the half-duration of a single pulse. 
According to the theoretical prediction\cite{Levitov:1996ie,Ivanov:1997wz} such a voltage pulse with a quantized flux, $\int_0^{\cal T} dt V  = h/e$, excites one electron out of the Fermi sea and no accompanying electron-hole pairs are excited. 
The recent ingenious experiment has successfully confirmed this theoretical prediction and unambiguously demonstrated a single-particle nature of excited quasiparticles  termed levitons.\cite{Dubois:2013dv}


First, we consider the case when levitons are excited in a ballistic conductor, $R=0$.
Substituting $V^L(t)$ into Eq.~(\ref{10}) we find a dc heat current, $J _{Q} ^{L} = \epsilon _{0}/{\cal T}$ (superscript $"L"$ stands for the flux of levitons).
Here $ \epsilon _{0} = \hbar/(2 \Gamma)$ is the mean energy of a leviton.\cite{Keeling:2006hq,Moskalets:2009dk,Dubois:2013fs}. 
Importantly the total heat power released in the system is two times bigger, $2 J _{Q} ^{L}$. 
The half of it is released in the contact where the levitons were exited from and the other half is carried by the stream of levitons to another contact.  
The heat noise, Eq.~(\ref{11c}), reads 
\begin{eqnarray}
{\cal S} _{Q} ^{L} =
\frac{ \hbar \Omega   }{ {\cal T} }
\left\{  
k_B\theta
+
\sum\limits_{q=1}^\infty  
q \hbar \Omega
\coth\left(  \frac{ q \hbar \Omega }{2k_B\theta } \right) 
e ^{ - q \frac{\hbar \Omega  }{  \epsilon_0 } }
\right\}
. 
\label{13} 
\end{eqnarray}
Given the Fourier coefficients of the applied potential, $\left |  eV ^{L} _{0}\right | ^{2}= (\hbar \Omega) ^{2}$, $\left |  eV ^{L} _{q\ne 0}\right | ^{2} = ( \hbar \Omega) ^{2} \exp( - q \hbar \Omega/ \epsilon _{0})$, and the equilibrium electrical noise, ${\cal S} _{I} ^{eq}( q\Omega) = (e ^{2}/h) q\hbar \Omega \coth\left( q\hbar \Omega/ [2k_B\theta] \right)$,   
we see that the equation above is nothing but Eq.~(\ref{main}). 

It is instructive to look at the noise to current ratio (or,  alternatively, at the heat Fano factor\cite{Sanchez:2013jx}), ${\cal E} ^{L} = {\cal S} _{Q} ^{L}/J _{Q} ^{L}$, which provides a characteristic energy.  
At high temperature, $k_B\theta \gg \epsilon _{0}$, the equation above gives ${\cal E} ^{L} = 2k_B\theta$.
This means that the thermal fluctuations are the dominant source of a heat noise. 

At low temperatures, $k_B\theta \ll \hbar \Omega$, however, we find ${\cal E} ^{L} = \epsilon _{0}$. 
This result reveals another source of a heat noise. 
To clarify it we write explicitly, ${\cal S} _{Q} ^{L} = \epsilon _{0} ^{2}/{\cal T}$. 
The quantity $\epsilon _{0} ^{2}$ is interpreted as the variance of leviton's energy, which agrees well with calculations based on the energy distribution function for single-electron excitations having Lorentzian time profile.\cite{Keeling:2006hq,Moskalets:2014ea}
Since the stream is regular, i.e., the number of electrons in the stream does not fluctuate (the electrical noise is zero), the intrinsic fluctuations of energy of levitons are the only source of heat fluctuations. 
The heat Fano factor in this case is just the ratio of the variance to mean energy of levitons, ${\cal E} ^{L} \equiv {\cal S} _{Q} ^{L}/J _{Q} ^{L} = \epsilon _{0} ^{2}/\epsilon _{0}  = \epsilon _{0}$.

Now we switch on backscattering, $R \ne 0$. 
We use the phase factor for levitons, $\Upsilon^L(t) = e^{2\pi i \frac{t}{\cal T} } (t + i \Gamma)/(t - i \Gamma)$, to calculate necessary Fourier coefficients and find  at low temperature, $k_B\theta \ll \hbar \Omega$, from Eqs.~(\ref{10}) and (\ref{11c}) heat fluctuations  ${\cal S} _{Q} ^{L} = T  (1+2R)\epsilon _{0} ^{2}/{\cal T}$\cite{note4,Battista:2014tj} and dissipated heat $J _{Q} ^{L} = T \epsilon _{0}/{\cal T}$. 
Thus the heat Fano factor becomes enhanced,  ${\cal E} ^{L} = (1+2R)\epsilon _{0}$. 
%
Similarly one can find that in the case of a  cosinusoidal driving voltage, $V _{1}(t) = V _{11} \cos( \Omega t)$, the heat Fano factor increases with increasing of a reflection coefficient too. 
At zero temperature it is, 
\begin{eqnarray}
{\cal E}  =  \left( 1 + \xi R \right)  \hbar \Omega ,
\label{hnenh}
\end{eqnarray}
where $\xi = -1 + (2/x)^2 \sum _{q=1} ^{\infty} q ^{3} J _{q} ^{2}(x)$ with $x = eV _{11}/\hbar \Omega$ and $J _{q}$ being the Bessel function of the first kind of the $q$th order. 

For comparison let us consider the stream of single particles (electrons and holes) emitted adiabatically by an on-demand coherent single-electron emitter\cite{Feve:2007jx} made of a quantum capacitor\cite{Buttiker:1993wh,Gabelli:2006eg} side attached to a chiral electronic waveguide. 
The capacitor is driven by an ac potential applied to a top gate.  
The stream of particles is split at the QPC, as shown in Fig.~\ref{hallbarQPC} but with $V=0$ and a capacitor being attached at  the left upper piece of a chiral waveguide. 
Since these particles have Lorentzian profile, their mean energy and the variance of energy are the same as for levitons. 
However, it turns out that the heat Fano factor (at zero temperature), ${\cal E} = \epsilon _{0}$,  is independent of the reflection coefficient $R$. 
The difference with the case of levitons comes from the fact that after reflection at the QPC particles arrive at the contacts, where the potential (driving a capacitor) is not applied to. 

In conclusion, 
the Floquet scattering matrix theory of heat fluctuations in driven coherent multi-terminal conductors is developed. 
The intrinsic fluctuations of energy of dynamically excited electrons are identified as the  fundamental source of the low-frequency heat noise, which does not show itself in the low-frequency electrical noise. 
In the linear response regime or at high temperatures the heat noise is equilibrium. 
Otherwise the non-equilibrium nature of heat noise manifests itself in the presence of scattering back to original contact with oscillating potential. 



\begin{thebibliography}{11}
\bibitem{note3}
The general equations (\ref{08a}) and (\ref{08b}) give a heat flux and its fluctuations also for unequal temperatures of the reservoirs.
\bibitem{note1}
If the thermal resistance of the contacts is not negligible, the Joule heat can be  partitioned asymmetrically, see, e.g., Refs.~\onlinecite{Itskovich:1987tu,Rokni:1995tz}.
\end{thebibliography}
\end{document}